\documentclass[12pt]{article}
\usepackage{amsmath}
\usepackage{epsfig}
\usepackage{pstricks,pst-plot}
\usepackage{amstext,amssymb,amsfonts}
\usepackage[all]{xy}

\advance\voffset by -2.0cm
\advance\hoffset by -1.25cm
\def\tr{\,{\rm tr}\, }

\def\be{\begin{equation}}
\def\ee{\end{equation}}
\def\ba{\begin{eqnarray}}
\def\ea{\end{eqnarray}}

\newcommand{\p}{\partial}
\newcommand{\bp}{\bar\partial}

\newcommand{\eg}{{\it e.g.~}}
\newcommand{\ie}{{\it i.e.~}}

\newcommand{\OOO}{{\cal O}}

\newcommand{\phicross}
{
\psline(-.05,-.05)(.05,.05)
\psline(-.05,.05)(.05,-.05)
\rput(0.15,0.15){$\phi$}
}
\newcommand{\chicross}
{
\psline(-.05,-.05)(.05,.05)
\psline(-.05,.05)(.05,-.05)
\rput(0.5,0.15){$\chi(0)$}
}

\begin{document}

\vspace*{-1.5cm}
\thispagestyle{empty}
\begin{flushright}
\end{flushright}
\vspace*{2.5cm}

\begin{center}
{\Large 
{\bf Brane backreactions and the Fischler-Susskind
\linebreak\linebreak
 mechanism in  conformal field theory}} 
\vspace{2.5cm}

{\large Christoph A.\ Keller}%
\footnote{{\tt E-mail: kellerc@itp.phys.ethz.ch}} 
\vspace*{0.5cm}

Institut f{\"u}r Theoretische Physik, ETH Z{\"u}rich\\
CH-8093 Z{\"u}rich, Switzerland\\
\vspace*{3cm}

{\bf Abstract}

\end{center}
The backreaction of D-branes on closed string moduli is studied in
perturbed conformal field theory.  
To this end we analyse the divergences in the modular integral of the
annulus diagram. 
By the Fischler-Susskind mechanism, these divergences  
lead to additional terms in the bulk renormalisation group
equations. We derive explicit expressions for these
backreaction terms, and follow the resulting 
renormalisation group flow in  
several examples, finding agreement with geometric expectations.

\noindent 
\newpage
\renewcommand{\theequation}{\arabic{section}.\arabic{equation}}


\section{Introduction}
A lot of recent work in string theory has dealt with the question of
moduli and moduli stabilisation in realistic compactifications. In
such setups two kinds of moduli appear. Closed string moduli
correspond to deformations of the bulk theory, \ie in geometric
language to deformations of the compactification manifold. Open string
moduli on the other hand correspond to deformations of the branes of
the configuration.

\noindent String compactification can also be considered in the
framework of
two dimensional conformal field theory. The compactification
is then no longer given by a Calabi-Yau manifold, but by a
worldsheet CFT of the correct central charge. The branes of such a
configuration are described by conformal boundary conditions.
Closed string moduli are given by exactly
marginal bulk operators, open string moduli by exactly marginal
boundary operators, and the theory is deformed by inserting such
integrated operators in the correlators.
Arguably the CFT point of view is more fundamental, as it
includes all $\alpha'$ corrections. On the other hand
only for very few geometric configurations the
corresponding worldsheet CFT is known explicitly.

\noindent To determine the moduli space of the theory, one needs to
find all exactly marginal
operators. A marginal operator is exactly marginal if it
remains marginal in the perturbed theory, or, to put it another way,
if it does not run under the renormalisation group flow. Criteria for
this have been worked out for bulk \cite{Chaudhuri:1988qb} and
boundary operators \cite{Recknagel:1998ih}. 

\noindent More recently, \cite{Fredenhagen:2006dn} considered the
interplay between bulk and boundary operators. In particular,   
renormalisation group flow equations were derived which describe the
effects of bulk perturbations on the boundary. These
equations describe how the open string moduli space changes as
bulk perturbations are turned on (see also \cite{Fredenhagen:2007rx}
for a discussion of this question).
More generally, they show that 
the boundary conditions flow to a fixed point which is compatible
with the new, perturbed bulk theory. The bulk theory, however, 
remains fixed and is not affected by the boundary conditions --- the
brane does not backreact on the bulk.

\noindent The aim of this paper is to extend the RG
equations of \cite{Fredenhagen:2006dn} to include the backreaction of 
branes on the bulk theory. The idea for the underlying mechanism goes
back to \cite{Cremmer:1973ig,Clavelli:1973uk,
  Fischler:1986ci,Fischler:1986tb,Callan:1986bc}: 
in string theory, to calculate amplitudes one considers not only the
disk diagram, but also diagrams of higher genus. The
total amplitude is obtained by summing over 
all topologies and integrating over the moduli of the conformal
structure of the diagrams. This integration can lead to new
divergences at the boundary 
of the moduli space $\mathcal{M}$, \ie when the surface
degenerates. More precisely, the spectrum of the 
theory may 
contain tadpoles, \ie massless modes, which give logarithmic
divergences when integrated over $\mathcal{M}$. According to
\cite{Fischler:1986ci,Fischler:1986tb}, these 
can be absorbed by a suitable shift of the coupling
constants in lower genus diagrams, thus contributing to the RG flow of
the bulk couplings. Since the nature of the tadpoles depends on the boundary
condition that is imposed, this describes the backreaction of the brane
on the bulk. 

\noindent We show  that this prescription
works for the annulus diagram, \ie that the tadpole divergences can be
compensated by local
counterterms on the disk diagram, leading to additional terms in
the bulk RG equations of \cite{Fredenhagen:2006dn}. The brane
backreaction can thus be  
incorporated quite naturally in the language of renormalisation group
flows. 

\noindent The RG equations so obtained can be used to study various
examples. In many cases, we know already from geometric
considerations how the brane should deform the bulk theory, so
that we can compare our results. For instance, we expect that
a D1-brane wrapping a circle should
shrink its radius. This is confirmed by the RG
analysis. In other, more complicated examples we also find
agreement between the RG analysis and geometric expectations or
supergravity calculations. 
 
\noindent This paper is organised as follows. In section~2 we first
rederive the bulk-boundary RG equations of \cite{Fredenhagen:2006dn}
using a different regularisation scheme which is more suitable for
further analysis. We then derive the backreaction term to first order
in the string coupling constant $g_s$ by
analysing divergences of the  annulus diagram. In section~3 we apply
the extended RG equations to 
 the free boson and WZW models. Section~4 discusses
bosonic string theory in flat space and its relation to supergravity
solutions. Finally, section~5 contains our conclusions.


\section{Renormalisation group equations}
\subsection{Dimensional regularisation on the disk}
Let us first derive the renormalisation group equations on the disk
\cite{Cardy, Fredenhagen:2006dn}. Consider the partition function
$\langle e^{-S}\rangle$, where $S$ 
is the perturbed action,
\be
S=S^\ast-\Delta S= S^\ast -\sum_i\lambda_i \,
\ell^{h_{\phi_i}-2}\int\phi_i(z)\, d^2z
-\sum_j \mu_j \, \ell^{h_{\psi_j}-1} \int\psi_j(x)\, dx\ .
\ee
We have introduced the length scale $\ell$ to keep the coupling
constants dimensionless. Expanding $\langle e^{-S^\ast+\Delta S}\rangle$ in
powers of $\lambda_i$ and $\mu_j$ gives terms of the form
\begin{multline}
\frac{\lambda_1^{l_1}\cdots\mu_1^{m_1}\cdots}{l_1!\cdots m_1!\cdots}
\prod_i \ell^{(h_{\phi_i}-2)l_i}\prod_j \ell^{(h_{\psi_j}-1)m_j} \\
\times\int\langle\phi_1(z_1^1) \phi_1(z_2^1)\cdots\phi_2(z^2_1) \cdots
\psi_1(x_1^1)\cdots\rangle \prod d^2z_k^i \prod dx_k^j \ .\label{int}
\end{multline}
Here the bulk fields $\phi_i$ are integrated over the entire disk, and the
boundary fields $\psi_j$ over its boundary.
The disk has the conformal symmetry group $SU(1,1)$.
The integration measure $d\mu$ must transform with conformal weight
$(-1,-1)$ under such transformations, so that integrals of marginal
$(1,1)$ fields $\int d\mu \phi_{(1,1)}$ are invariant. Clearly, $d^2z$
satisfies this property. Since we can use $SU(1,1)$ to map any
point to $0$, it follows that up to a constant factor
this is the only possible measure. 

\noindent Because of the symmetry group,
the integrals in (\ref{int}) are infinite. To render them finite, we
use $SU(1,1)$ to fix the position of  
one bulk and one boundary insertion. Alternatively, we can (formally) divide by
the volume of $SU(1,1)$. 

\noindent The terms (\ref{int}) are still infinite, since
the integrand diverges when fields come close together. More
precisely, three different situations can cause divergences:
when two bulk fields come close to each other, when two boundary fields come
close to each other, or when a bulk field comes close to the
boundary. These three 
situations will lead to the three different terms in the RG equations
(\ref{bulkflow}) and (\ref{bounflow}) below.
We thus have to introduce a scheme to regularise the divergences.
One such scheme \cite{Cardy, Fredenhagen:2006dn} is to
cut out small disks of radius $a$ around all operators. 

\noindent Instead we will 
use a scheme which resembles dimensional
regularisation. To evaluate diverging integrals, we change the
conformal dimension of the fields involved to such values that the
integral converges, and evaluate the original integral by
analytic continuation. One motivation for using 
this scheme comes from the spacetime interpretation of the 
divergences that will show up in the modular integrals: they can be
interpreted as infrared divergences due to massless modes,
so that a natural regularisation is to introduce a small mass term. In
the worldsheet theory, this corresponds to a shift of the
conformal dimension of the field.
From a more technical point of view, it is
favourable to keep conformal covariance of all expressions, which is
destroyed if we cut out small disks.

\noindent Let us shift the conformal weight
of boundary fields as $h_\psi \mapsto h_\psi - \epsilon$, and that of
bulk fields as $h_\phi \mapsto h_\phi -2\epsilon$.\footnote{Note that
  for bulk fields in a theory
with boundary $h=h_L+h_R$.} As an example for how the scheme works,
consider two marginal 
bulk fields $\phi_i, \phi_j$ that come close to each other to produce
another marginal field $\phi_k$,
\be
\lambda_i\ell^{-2\epsilon}\lambda_j\ell^{-2\epsilon}\phi_i(z)\phi_j(0) \sim
\lambda_i\lambda_j\ell^{-4\epsilon}\frac{\phi_k(0)C_{ijk}}{|z|^{h_i+h_j-h_k}}
=  
\lambda_i\lambda_j\ell^{-4\epsilon}\phi_k(0)C_{ijk}|z|^{-2+2\epsilon}
\ .
\label{2ptOPE}
\ee
For simplicity, we have fixed the position of $\phi_j$ to 0.
We perform the $d^2 z$ integral up to some IR cutoff $L$ to obtain  
\be
\lambda_i\lambda_j\ell^{-2\epsilon}\phi_k(0)\,2\pi C_{ijk}
\frac{\ell^{-2\epsilon}}{2\epsilon}L^{2\epsilon}\ . \label{epsilonpole}
\ee
We have pulled out a factor $\ell^{-2\epsilon}$ which will be
absorbed in the shift of $\lambda_k$ (see (\ref{shift})). The second
factor $\ell^{-2\epsilon}$ gives
\be
\frac{\ell^{-2\epsilon}}{2\epsilon}L^{2\epsilon}= \frac{1}{2\epsilon}
- \log \ell + \log L + \OOO(\epsilon)\ .
\ee
In the limit $\epsilon\rightarrow\infty$, only the second term gives a
dependence on $\ell$ which contributes to the RG flow. 
We see that the regularisation scheme has  introduced an
implicit dependence of the 
integral on $\ell$. As $\langle e^{-S}\rangle$ must be independent of
$\ell$, we must compensate a shift in $\log\ell$ by shifting $\lambda_i$
and $\mu_j$. A combinatorial analysis shows that the shift needed
is 
\be
\lambda_k\ell^{-2\epsilon} \mapsto \lambda_k\ell^{-2\epsilon} + 
\lambda_i\lambda_j\ell^{-2\epsilon}\pi C_{ijk}\cdot \log \ell\
. \label{shift} 
\ee
In a similar way, we treat the other types of divergences. The
resulting renormalisation group equations are \cite{Fredenhagen:2006dn}
\begin{eqnarray}
\dot\lambda_k &=& (2-h_{\phi_k})\lambda_k +
\pi C_{ijk}\, \lambda_i\lambda_j + {\cal O}(\lambda^3)\ , \label{bulkflow}\\
\dot\mu_k &=& (1-h_{\psi_k})\mu_k + \frac{1}{2}\, B_{ik}\, \lambda_i  
+ D_{ijk}\, \mu_i\mu_j + {\cal O}(\mu\lambda, \mu^3, \lambda^2)\ ,
\label{bounflow}
\end{eqnarray}
where the dot indicates a derivative with respect to the flow
parameter $t = \log \ell$. 
To obtain higher order terms in $\mu$ and $\lambda$, one
would have to analyse the 
situation when three or more fields come close to each other. 
In the following, we shall never consider such terms.

\subsection{Higher genus: general strategy }
To calculate amplitudes in 
string theory, we have to take into account higher genus diagrams as well.
For simplicity assume that there is only
one type of field $\phi$ in our theory. As before, a string
amplitude $F$ can be expanded in powers of $\lambda$,
$ F=\sum_n \lambda^n F_n$ .
Each term $F_n$ itself contains contributions from all topologically
different diagrams with $n$ insertions of $\phi$. Moreover, for a
given topology we must integrate over all conformal structures,
parametrised by modular parameters $t_i$. In full, 
\be
F_n = \sum_k g_s^{\chi_k} \int_{\mathcal{M}_k} d t_i F_n^{k}(t_i)\ ,
\ee
where $g_s$ is the string coupling constant and $\chi_k$ is the Euler
characteristic of the diagram $F^k$.
Integration over the moduli space $\mathcal{M}_k$ leads to new divergences
due to marginal and relevant modes in the spectrum of the theory. The
divergences have to be regularised, and we 
must try to compensate for them by introducing counterterms on diagrams
of lower genus. These $\ell$-dependent terms then give the
the backreaction terms in the bulk RG equations.

\subsection{The annulus diagram} \label{ssAnnulus}

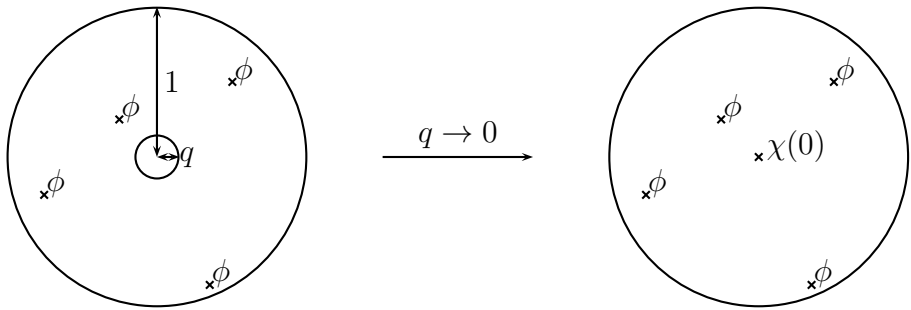
\begin{figure}
\begin{center}

\begin{pspicture}(0,-2.5)(10,2.5)

\psellipse(1,0)(2,2)
\psellipse(1,0)(0.3,0.3) 
\psline{<->}(1,0)(1.3,0)
\rput(1.4,0){$q$}
\psline{<->}(1,0)(1,2)
\rput(1.2,1){$1$}

\rput(0.5,0.5){\phicross}
\rput(-0.5,-0.5){\phicross}
\rput(2,1){\phicross}
\rput(1.7,-1.7){\phicross}

\psline{->}(4,0)(6,0)
\rput(5,.3){$q\rightarrow 0$}

\psellipse(9,0)(2,2)
\rput(9,0){\chicross}

\rput(8,0){
\rput(0.5,0.5){\phicross}
\rput(-0.5,-0.5){\phicross}
\rput(2,1){\phicross}
\rput(1.7,-1.7){\phicross}
}

\end{pspicture}

\end{center}
\caption{Divergences of the annulus diagram}
\label{fig1}
\end{figure}

We will now calculate the backreaction terms caused by the annulus
diagram $A_n=F^1_n$.  
The annulus has a single real modular  
parameter $q$, its inner radius. The integral over $q$ produces a
divergence for $q\rightarrow 0$. 
In this case there is an intuitive way to see how the counterterm on
the disk arises, as shown in figure \ref{fig1}:
the divergent part of the annulus
diagram with $n$ integrated insertions corresponds to a 
disk diagram with an additional field $\chi(0)$ inserted. A shift 
$\lambda\mapsto\lambda+\delta\lambda$ on the disk diagram
$D_{n+1}=F^0_{n+1}$ can thus
compensate the divergence. The corresponding term is of order $g_s$.

\noindent Although we will only calculate the 
term of order $g_s\lambda^0$, some comments on terms of higher order
in $\lambda$ are
necessary. The analysis on the disk showed that $\lambda^2$ terms are 
produced by two fields  
approaching each other, and that higher order terms 
appear when $n$ fields come close together.
In the situation here, higher order corrections arise when 
additional fields move 
close to the new field produced on the disk or to the boundary of the
annulus. If for instance a single $\phi$ moves 
close to the centre of the annulus $A_n$, the divergence can be
compensated by the disk diagram $D_{n}$, which produces
a contribution of order $g_s\lambda$.
As we are only interested in the lowest order correction, we
can thus subtract divergences which arise from fields moving 
close to each other or to the boundary.

\noindent Note that the symmetry group of the annulus is only $U(1)$
--- we can 
fix the position of one boundary insertion, or alternatively we
can divide the amplitude by $2\pi$. This also means that unlike
on the disk, the conformal
symmetry no longer uniquely fixes the integration measure. 
Nevertheless, the correct measure is still $d^2 z$, see \eg
\cite{Green:1987mn}.

\noindent
For a given radius $q$, the integrated $n$-point amplitude of the annulus
is given by
\be
A_n(q) = \frac{1}{\pi} \prod_{i=1}^n\int_1^q d^2z_i
\langle\langle B\vert\vert  
\phi(z_1)\ldots \phi(z_n) 
q^{L_0+\bar L_0 -2} \vert\vert B\rangle\rangle\ . \label{A_n}
\ee
For simplicity, we have only included one type of marginal field $\phi$. 
As usual, $\langle\langle B\vert\vert$ is the boundary state at the
outer radius 1. 
To obtain the boundary state at the inner radius, we transport
$\vert\vert B\rangle\rangle$
to the inner radius $q$ using the propagator $\pi^{-1}q^{L_0+\bar L_0 -2}$,
whose normalisation is fixed by the construction of the boundary states.
By inserting a complete set of states, we expand the boundary state
in a sum of fields inserted at the point 0. The action of the
propagator then gives 
\be
\pi^{-1}q^{L_0+\bar L_0-2}||B\rangle\rangle = \pi^{-1}\sum_i
q^{h_i+\bar h_i-2}|\phi_i\rangle \langle\phi_i||B\rangle\rangle\ 
\label{Btransported}\ .
\ee
Here $\langle\phi_i||B\rangle\rangle$ is the disk one-point function
with $\phi_i$ sitting at the point 0.
Integrating (\ref{A_n}) over its moduli space using the
measure $q^{-1} dq$, we see from 
(\ref{Btransported}) that divergences arise for $q\rightarrow 0$
for all fields with $h_i = \bar h_i \leq 1$. 
In a supersymmetric setup, we expect no relevant, \ie tachyonic
fields. In the bosonic theories we will consider, the only such field is
usually the vacuum $h=\bar h =0$. The vacuum only changes overall
normalisations, so that we will ignore it in what follows.
The only divergences are then due to marginal fields 
$h_i = \bar h_i = 1-\epsilon$.
Their contribution is 
\be
||B(q)\rangle\rangle\simeq 
 \frac{q^{-2\epsilon}}{\pi} 
\sum_i\langle\phi_i||B\rangle\rangle \phi_i(0)\ .  \label{bdstmarg}
\ee
For the moment, let us assume that there are no integrated bulk insertions.
The integral of (\ref{bdstmarg}) over moduli space converges if
$\epsilon < 0$, and we  
will use its analytic continuation,
\be
\int_0^1\! q^{-1}dq\, ||B(q)\rangle\rangle =
-\frac{1}{\pi}\frac{1}{2\epsilon}
\sum_i\langle\phi_i||B\rangle\rangle \phi_i(0)\ . \label{pole}
\ee
The pole in $\epsilon$ will then contribute to the
RG equations as in (\ref{epsilonpole}). 

\noindent  If the diagram contains integrated
bulk insertions, the comparison is a bit more subtle: in the disk
diagram, the additional bulk insertions are integrated over the entire
disk, whereas on the annulus they are only integrated up
to the inner radius $q$. The divergent contribution of the tadpole,
however, comes from the limit $q\rightarrow 0$. We can thus
concentrate on annulus diagrams where $q < |\epsilon|$. Indeed,  
\be
\int_{|\epsilon|}^1 dq q^{-1-2\epsilon} =
-\frac{1}{2\epsilon}(1-e^{-2\epsilon\ln{|\epsilon|}}) 
 = \OOO(\ln|\epsilon|)
\ee
is only a subleading contribution compared to (\ref{pole}).
We claim then that to lowest order in $\lambda$ we can rewrite
the annular integral as 
\be
\int^{|\epsilon|}_0 dq \int_q^1 d^2z_i\langle\ldots\rangle
= \int_{|\epsilon|}^1 d^2z_i \int^{|\epsilon|}_0 dq \langle\ldots\rangle
+ \OOO(\epsilon^2)\ .
\ee
This holds because we can estimate the contribution of the fields
$\phi$ integrated over the small disk of radius $|\epsilon|$: since we
only calculate the lowest order term in 
$\lambda$, we subtract all singular terms in $\phi$. The remaining
expression is
then bounded by some constant $B$, and we can estimate its
contribution as $\leq  
\pi\epsilon^2B$. 
A similar argument shows that we can cut out the same
small disk in the disk diagram without changing the result. This shows
that we can compare annulus diagrams with disk diagrams even if they
contain integrated insertions.

\noindent So far, the
fields $\phi_i$ introduced by the tadpoles are 
inserted at the point $z=0$. In 
order to be able to compensate them with a disk diagram, we
need to rewrite them as integrated 
insertions. To do this, we use the fact that the disk has a larger
symmetry group than the annulus. Consider the disk diagram with $n$
integrated fields $\phi(z_i)$ and one additional field $\chi(z)$, each
of them marginal.  
We can use part of the symmetry group $SU(1,1)$ to fix the 
position of $\chi$ to 0. In particular, for each $z$ 
choose $f_z \in SU(1,1)$ such that $f_z(z)=0$. Defining $\hat z_i
= f_z(z_i)$, conformal covariance tells us that the $z_i$ integral 
changes as 
\be
\int d^2z_i\phi(z_i) \rightarrow \int d^2\hat z_i \left|
\frac{\partial z_i}{\partial \hat z_i}\right|^{-2\epsilon} \phi(\hat z_i)
= \int d^2\hat z_i \phi(\hat z_i) + \OOO(\epsilon)\ . \label{Jacobian}
\ee
Up to terms of order $\epsilon$, the resulting integral is thus independent
 of $z$, and the additional field $\chi(z)$ is fixed at the position
 $z=0$.
Formally, we can write this manipulation as
\be
\frac{1}{|SU(1,1)|}\int d^2z \int d^2z_i
\langle\chi(z)\phi(z_1)\ldots \rangle = \frac{1}{|U(1)|} \int
d^2\hat z_i \langle\chi(0)\phi(\hat z_1)\ldots \rangle +
\OOO(\epsilon)\ , \label{symGroup}
\ee
where $|G|$ denotes the volume of the respective symmetry group. On
the right hand side of (\ref{symGroup}), we divide by $|U(1)|$ because
we still have not fixed the entire symmetry:
after choosing $f_z$, we can always rotate the disk around its centre. 
This remaining $U(1)$ symmetry however is exactly
the symmetry group of the annulus, so that the right hand side of
(\ref{symGroup}) is the standard annulus diagram with one fixed insertion.

\noindent  The upshot of this
analysis is that the divergent part of  
$A_n$ has the same form as $D_{n+1}$, so that it can be compensated 
by introducing a counterterm on the disk diagram.
As before, we need to
split off a factor $\ell^{-2\epsilon}$ to be included in
$\lambda$. The annulus contribution to the disk diagram is thus 
\be
-\ell^{-2\epsilon} \frac{g_s}{\pi}
\frac{\ell^{2\epsilon}}{2\epsilon}\langle\phi_i 
||B\rangle\rangle \int\! d^2z\, \phi_i(z) = -\ell^{-2\epsilon}
\frac{g_s}{\pi} \left(\frac{1}{2\epsilon} + 
\log \ell + \OOO(\epsilon)\right)\langle\phi_i 
||B\rangle\rangle \int\! d^2z\, \phi_i(z)
\ee
for each marginal field $\phi_i$.
The usual combinatorial analysis shows that this can be
compensated by shifting the coupling constant $\lambda_i$.

\noindent
Putting everything together we obtain the modified bulk RG equations
\be
\dot\lambda_k = (2-h_{\phi_k})\lambda_k + \frac{g_s}{\pi}
\langle\phi_k||B\rangle\rangle  +
\pi C_{ijk}\, \lambda_i\lambda_j + {\cal O}
(g_s\lambda,\lambda^3,g_s^2)\ . \label{g1bulkflow}  
\ee


\section{WZW models and the free boson}
\setcounter{equation}{0}
We now apply equation (\ref{g1bulkflow}) to some
examples. First we consider the free boson compactified on a
circle, subject to Neumann or Dirichlet boundary conditions.
Then we turn to Wess-Zumino-Witten
models based on compact Lie groups. These models and their
boundary states are very well understood and can be interpreted
geometrically.  We can thus check RG flow results against geometric
expectations. 

\subsection{The free boson on a circle}
Let $X(z,\bar z)$ be the free boson compactified on a circle of radius
$R$, $X \sim X + 2\pi R$. Its action is given by
\be
S = \frac{1}{2\pi}\int\! d^2z\, \partial X \bar\partial X\ .
\ee
Neumann and Dirichlet boundary conditions are given by 
identifying on the real axis $z=\bar z$
$$
\partial X = \bar\partial X \ \textrm{(Neumann)}\qquad
\textrm{and} \qquad  \partial X = -\bar\partial X \
\textrm{(Dirichlet).} 
$$
As usual, we can switch to the closed string picture by mapping the
upper half-plane to the disk. The boundary condition is then described by
the boundary states $||N\rangle\rangle$ and $||D\rangle\rangle$,
respectively. 

\noindent The ground states of the system are parametrised by momentum
and winding numbers $n,w \in \mathbb{Z}$ such that
\be
(p_L,p_R)= \left(\frac{n}{2R}+wR,\frac{n}{2R}-wR\right)\ , \label{spec}
\ee
with conformal weight given by
$(\frac{1}{2}p_L^2,\frac{1}{2}p_R^2)$. 
At a generic radius $R$, the only marginal operator is $\partial X
\bar \partial X$. Its one-point function is given by
\be
\langle \partial X \bar \partial X||N\rangle\rangle =1\  \qquad
\textrm{and} \qquad 
\langle \partial X \bar \partial X||D\rangle\rangle =-1\ .
\ee
We will also have to deal with the relevant fields that are present
in theory.

\noindent Let us analyse the Neumann case first. The one-point
function 
vanishes unless $p_L= -p_R$, \ie $n=0$, so that only pure winding
modes couple. If we take $R$ big enough, (\ref{spec}) shows that
all these modes become irrelevant. It is thus sufficient to only
consider the perturbation by $\partial X \bar \partial X$, 
\be
S = \frac{1}{2\pi} \int\!  d^2z\, \partial X \bar\partial X\ -
\lambda\int \! d^2z\, \partial X \bar\partial X\ . \label{pertaction}
\ee
We see that (\ref{g1bulkflow}) yields $\dot\lambda = g_s/\pi>0$. An
increase 
in $\lambda$ means that the circle shrinks, as can be seen from
(\ref{pertaction}): to maintain the correct normalisation of the
action, we have to 
introduce rescaled fields $X' = \sqrt{1-2\pi\lambda}X$, which satisfy
$X'\sim X'+2\pi R'=X' +2\pi R\sqrt{1-2\pi\lambda}$. 

\noindent This shows that a Neumann brane that wraps
the circle shrinks its radius. Similar reasoning
shows that the D0 brane given by $||D\rangle\rangle$  
increases the radius of the circle.

\noindent When $R$ becomes of the order
of the self-dual radius $R_0=1/\sqrt{2}$,  
new relevant and marginal fields appear, and the above analysis breaks
down. To analyse this case, we will use the fact that the free boson
at the self-dual radius is equivalent to the $SU(2)$
Wess-Zumino-Witten-model at 
level 1. We therefore turn our attention to WZW-models.

\subsection{Renormalisation group flows in general WZW models}
\label{ssWZW}
Wess-Zumino-Witten models are often described as $\sigma$-models on
a group manifold of a Lie group $G$ \cite{Witten:1983ar}. A different,
more algebraic approach is to define them via their operator content
and correlation functions. For the moment, we will use this more abstract
formulation, before changing to a more geometric picture in the next
section.

\noindent
The currents of the WZW model of a Lie group $G$ at level $k$
correspond to elements of the Lie algebra $\mathfrak{g}$ of $G$ 
and satisfy the operator product expansion
\be
J^a(z)J^b(w) \sim \frac{k\delta^{ab}}{(z-w)^2} +
if^{ab}_{\phantom{ab}c}\frac{J^c(w)}{(z-w)} \ ,
\ee
where $f^{ab}_{\phantom{ab}c}$ are the structure constants of
$\mathfrak{g}$. The marginal fields of the theory are given by all
possible combinations $J^a \bar J^b$ of left-moving and right-moving
currents. 
We consider branes that preserve the affine symmetry up to
conjugation by  $g \in G$
\cite{Gaberdiel:2001xm, Callan:1994ub, Polchinski:1994my}. In the
closed string picture this means that the boundary state
$||B\rangle\rangle$ has to satisfy the gluing condition
\be
(gJ^a_mg^{-1}+\bar J^a_{-m})||B\rangle\rangle=0\ ,
\ee
whereas in the open string picture the left and right moving
currents are identified at the boundary as
\be
gJ^a(z)g^{-1} = \bar J^a(\bar z) \quad \textrm{for } z=\bar z\ .
\ee
The one-point function is best evaluated in the open string picture
and gives \cite{FZ, Gaberdiel:1998fs}
\be
\langle (J^a \bar J^b)(u) \rangle_B = k\frac{\tr (J^a g
  J^b g^{-1})}  
{(u-\bar u)^2}= -k \frac{\tr (J^a g
  J^b g^{-1})}  
{|u-\bar u|^2} \ ,
\ee
so that $\langle J^a\bar J^b ||B\rangle\rangle = -k\tr (J^a g
  J^b g^{-1}) $.
Note that the currents are normalised such that $\tr (J^a J^b) =
\delta^{ab}$.
The orthonormal marginal fields are thus
\be
\phi_{ab}(z)=k^{-1}J^a \bar J^b\ . \label{norm}
\ee
Let us start from the model which is initially unperturbed. To lowest
order, (\ref{g1bulkflow}) gives then
\be
\dot \lambda_{ab} = -\frac{g_s}{\pi} \tr (J^a g
  J^b g^{-1}) \label{WZWgs1}
\ .
\ee
Higher order contributions in the bulk come from evaluating connected
$n$-point 
functions.  They
are given \cite{FZ, Gaberdiel:1998fs} by the product of traces 
$k\tr(J^{a_1}\ldots J^{a_n})k\tr(\bar J^{b_1}\ldots \bar J^{b_n})$,  
so that in the normalisation (\ref{norm}) they go as 
$k^{2-n}$. In the limit $k\rightarrow\infty$ they only give subleading
contributions. 

\noindent Let us make a side remark. We can choose an orthogonal basis
$J^a, a = 1,\ldots, r$ for 
the left moving currents, and a corresponding basis $\bar J^b := g^{-1}
J^b g, b = 1,\ldots, r$ for the right moving
currents. (\ref{WZWgs1}) then shows that only the fields
$\phi_{aa}$ are switched on. Note that these fields leave
the boundary conditions unchanged, as 
\be
[t^a,g\bar t^a g^{-1}]=[t^a,t^a]=0 \ ,
\ee
which means that all $B_{ik}$ in (\ref{bounflow}) vanish, so that no
boundary fields are switched on \cite{Fredenhagen:2006dn}. The brane
changes the bulk without inducing a backreaction on itself.

\subsection{Geometric interpretation of $SU(2)_k$}
To get a geometric picture of the brane backreaction, we switch to a
more geometric description of WZW models.
We will concentrate on $G=SU(2)$. We can write this theory as a
$\sigma$-model 
on the group manifold, using the parametrisation \cite{Hassan:1992gi} 
\be
g = e^{i(\theta+\tilde\theta)\sigma_2/2} e^{i\phi\sigma_1/2}
e^{-i(\theta-\tilde\theta)\sigma_2/2} \ ,
\ee
or explicitly 
\be
g=\left(\begin{array}{cc}
\cos\frac{\phi}{2}\cos\tilde\theta + i\sin\frac{\phi}{2}\sin\theta  &
\cos\frac{\phi}{2}\sin\tilde\theta + i\sin\frac{\phi}{2}\cos\theta \\
-\cos\frac{\phi}{2}\sin\tilde\theta + i\sin\frac{\phi}{2}\cos\theta &
\cos\frac{\phi}{2}\cos\tilde\theta - i\sin\frac{\phi}{2}\sin\theta
\end{array}\right)\ .
\ee
At level $k$ the action then becomes 
\be
S_0(\phi,\theta,\tilde \theta)= \frac{k}{2\pi}\int d^2z \left(\frac{1}{4}\
\bp\phi\p\phi + \sin^2\frac{\phi}{2}\ \bp\theta\p\theta + \cos^2
\frac{\phi}{2} \ \bp\tilde\theta\p\tilde\theta + \cos^2\frac{\phi}{2}
(\bp\theta\p\tilde\theta - \bp\tilde\theta\p\theta)\right)\ . \label{WZWaction}
\ee
For later use, we also derive explicit expressions for the currents $J
= -k \p g \ g^{-1}$ and $\bar J = k g^{-1}\bp g$, 
\begin{eqnarray*}
J^1&=&-k\frac{i}{\sqrt{2}}(\p\phi \cos(\tilde\theta+\theta)
-\p\theta\sin\phi\sin(\tilde\theta+\theta)
+\p\tilde\theta\sin\phi\sin(\tilde\theta+\theta) \\
J^2&=&-k\frac{i}{\sqrt{2}}(\p\theta(1-\cos\phi)+\p\tilde\theta(1+\cos\phi))
\\
J^3&=&-k\frac{i}{\sqrt{2}}(\p\phi \sin(\tilde\theta+\theta)
+\p\theta\sin\phi\cos(\tilde\theta+\theta)
-\p\tilde\theta\sin\phi\cos(\tilde\theta+\theta)
\end{eqnarray*}
and
\begin{eqnarray*}
\bar J^1&=&k\frac{i}{\sqrt{2}}(\bp\phi\cos(\tilde\theta-\theta) +
  \bp\theta\sin\phi\sin(\tilde\theta-\theta) +
  \bp\tilde\theta\sin\phi\sin(\tilde\theta-\theta)) \\
\bar J^2&=&k\frac{i}{\sqrt{2}}(\bp\theta(-1+\cos\phi)
  +\bp\tilde\theta(1+\cos\phi)) \\
\bar J^3&=&k\frac{i}{\sqrt{2}}(-\bp\phi\sin(\tilde\theta-\theta) +
  \bp\theta\sin\phi\cos(\tilde\theta-\theta) +
  \bp\tilde\theta\sin\phi\cos(\tilde\theta-\theta))\ .
\end{eqnarray*}

\noindent
The boundary states are given by
$||j,g\rangle\rangle$. For each gluing condition $g$ there are
$k+1$ possible branes, labelled by
$j=0,\frac{1}{2},\ldots,\frac{k}{2}$.
\cite{Alekseev:1998mc} gives a geometric interpretation for these branes
in terms of conjugacy 
classes: if $g$ is the identity $e$, then  
$||j,e\rangle\rangle$ is the $S^2$ 
that wraps the conjugacy class given by
\be
h\left(\begin{array}{cc}
e^{2\pi i j/k}&0\\
0&e^{-2\pi i j/k} \end{array}\right)h^{-1}\ .
\ee
In particular, for $j=0$ and $j=\frac{k}{2}$, the conjugacy class
collapses to a point and the brane describes a D0 brane sitting at the
point $e$ and $-e$, respectively. If the gluing
map is given by a general $g$, the position of the brane shifts
accordingly. 

\noindent
To go to the geometric limit, we fix $j$ and let
$k\rightarrow\infty$. Independent of $j$ the brane thus becomes
a D0 brane sitting at the point $g$.
Also, (\ref{WZWgs1}) shows that the flow induced depends only on
$g$. We can therefore suppress the index $j$ and parametrise 
the brane only by $g=g(\Phi, \Theta, \tilde\Theta)$. Note that we 
denote its position by capital letters $\Phi, \Theta, 
\tilde \Theta$, as opposed to small letters for the coordinates of the
manifold.

\noindent 
In the geometric limit the $SU(2)_k$ model corresponds to a
non-linear $\sigma$-model on $S^3$ with radius $r \sim \sqrt k$. 
We can read off the target space metric $G$ and the
field $B$ from the coefficients of the action. In the unperturbed case
(\ref{WZWaction}) this gives
\be
G_0=\left(\begin{array}{ccc}
k/4 &0&0\\
0 & k \sin^2 \frac{\phi}{2} & 0\\
0&0& k \cos^2\frac{\phi}{2} \end{array}\right)\ , \qquad
B_0=\left(\begin{array}{ccc}
0&0&0\\
0&0&k\cos^2\frac{\phi}{2}\\
0&-k\cos^2\frac{\phi}{2}&0 \end{array}\right)\ .
\ee

\subsection{Minimising the brane mass}
Let us now calculate the RG flow and try to interpret it.
(\ref{WZWgs1}) shows that the marginal fields $J^i\bar J^j$ are turned
on with the respective strength
\be
\dot \lambda_{ij}(\Phi,\Theta,\tilde\Theta)=
-\frac{g_s}{\pi}\tr(J^igJ^jg^{-1}) =: 
-\frac{g_s}{\pi} K_{ij}(\Phi,\Theta,\tilde\Theta) \ . \label{su2flow}
\ee
The coefficients $K_{ij}$ depend on the position of the brane and are
given by
\be
 K_{ij}=
2{\small \left(\begin{array}{ccc}
 \cos 2\tilde\Theta\cos^2\frac{\Phi}{2}+
\cos 2\Theta\sin^2\frac{\Phi}{2} & \sin(\Theta+\tilde\Theta)\sin\Phi &
\sin 2\Theta \sin^2\frac{\Phi}{2}- 
\sin 2\tilde\Theta\cos^2\frac{\Phi}{2} \\ 
-\sin(\Theta-\tilde\Theta)\sin\Phi & \cos \Phi &
\cos(\Theta-\tilde\Theta)\sin\Phi \\
\sin 2\Theta\sin^2\frac{\Phi}{2}+
\sin2\tilde\Theta\cos^2\frac{\Phi}{2}
&-\cos(\Theta+\tilde\Theta)\sin\Phi  & \cos 2\tilde\Theta\cos^2\frac{\Phi}{2}-
\cos 2\Theta\sin^2\frac{\Phi}{2} \end{array}\right)} \ . \label{Kij}
\ee
This flow has a nice geometric interpretation. The mass of a brane
is given by the value of the 
dilaton $\varphi$. Perturbing the metric of $S^3$ induces a non-constant
dilaton and so changes the mass of the
brane. \cite{Fredenhagen:2006dn} showed that in the case of an induced
boundary flow, the brane deformed in such a way as to minimise its
mass. We will show that a similar thing happens here: this time, the
brane remains at the same place, but it deforms the geometry in such a
way that its mass is minimised.

\noindent To show this, let us first find the change in geometry that
decreases the mass of the brane as much as possible.
The most general current-current deformation of the original theory is
\be
S=S_0-\alpha\int d^2z \sum_{i,j}a_{ij}J^i(z)\bar J^j(\bar z) \ , \label{gPer}
\ee
where the $a_{ij}$ are real coefficients. This gives a new metric
$G'(\phi,\theta,\tilde\theta)=G_0-\alpha G_1$ and a
new $B$-field.  
The new, nontrivial dilaton $\varphi$ can be
calculated by
\cite{Hassan:1992gi,Forste:2001gn}
\be
e^{-2\varphi_0}\sqrt{\det G_0}=e^{-2\varphi(\phi,\theta,\tilde\theta)}
\sqrt{\det G'(\phi,\theta,\tilde\theta)}\ . 
\ee
The mass of the brane at $g=g(\Phi,\Theta,\tilde\Theta)$ is 
proportional to $g_s^{-1}\sim
e^{-\varphi(\Phi,\Theta,\tilde\Theta)}$. 
We thus  want to maximise the increase of $\det G_0^{-1}G'$ at the
point $(\Phi, \Theta, \tilde\Theta)$. 
Its derivative is given by
\be
\p_\alpha \det(1-\alpha G^{-1}_0G_1)|_0 = -\tr G^{-1}_0G_1\ . 
\ee
A straightforward calculation then shows 
\be
\tr G^{-1}_0G_1(\Phi,\Theta,\Tilde\Theta) =
k\sum_{i,j}a_{ij}K_{ij}(\Phi,\Theta,\Tilde\Theta)\ , 
\ee
where $K_{ij}$ is the same expression as in (\ref{Kij}).
Introducing a Lagrange multiplier term
$\nu \sum_{i,j}a_{ij}^2$ shows
that the expression is extremised by 
$a_{ij}=-K_{ij}(\Phi,\Theta,\tilde\Theta)$. 
Comparing to (\ref{su2flow}) we find perfect agreement.

\vspace{0.5cm}
\noindent We can try to follow the flow further 
and describe the geometry of the deformed manifold. By
the symmetry of the problem, it is sufficient to
consider the brane sitting at $\theta = 0, \tilde\theta
= 0$\footnote{We could simply restrict to $g=e$, but $e$ is a
  coordinate singularity in the parametrisation chosen.} so that
\be
g = \left(\begin{array}{cc}
e^{i\Phi/2}&0\\
0&e^{-i\Phi/2}
\end{array}\right)\ .
\ee
(\ref{Kij}) then turns on the fields
\be
\lambda_{ij}=-2\,\frac{g_s}{\pi}\left(\begin{array}{ccc}
1 & 0 & 0 \\ 
0 & \cos \Phi &\sin\Phi \\
0 &-\sin\Phi  & \cos\Phi \end{array}\right) \ .
\ee
They change the metric $G_0$ by some expression
$2\,\frac{g_s}{\pi}G_1^\Phi(\phi,\theta,\tilde\theta)$. At the point of
the brane, $G_1^\Phi$ simplifies:  
\be
G_1^\Phi(\Phi,0,0)=\left(\begin{array}{ccc}
k^2/2 &0&0\\
0 & 2k^2 \sin^2 \frac{\Phi}{2} & 0\\
0&0& 2k^2 \cos^2\frac{\Phi}{2} \end{array}\right) = 2kG_0(\Phi,0,0) \ .
\ee
The effect of the backreaction is simply to rescale the original
metric. We can
continue to use our original reasoning even away from the 
point $t=0$ to obtain the differential equation  
\be
\dot G^\Phi(\Phi,0,0) \sim G_0(\Phi,0,0) \ . \label{Gdiff}
\ee
The geometric analysis only gives the direction of the flow,
so that we are free to choose the actual flow parameter.
Writing
\be
G^\Phi_{\mu\nu}(t)=G_{0\mu\nu}+4\frac{g_s}{\pi}ktG^{\Phi}_{1\mu\nu}\
 \label{Gdeform} 
\ee
we fix $t$ so that it
agrees with  the conformal field theory flow parameter at $t=0$.

\noindent Note that this analysis agrees with the observation in
section~\ref{ssWZW}, where 
we argued that in the limit $k\rightarrow\infty$, only the zero order
term is important, and that thus no new bulk fields are
turned on. This translates to the statement that (\ref{Gdiff}) remains
valid away from the starting point.

\noindent We can now try to
understand how the geometry of the $S^3$ changes as we start to flow,
and we can also try to estimate how far we should trust our analysis. 
Define a new flow parameter $t'= 4g_st/\pi$.
Take the metric $G^\Phi_{\mu\nu}(t')$ and
calculate the associated Ricci scalar $R(t')$. At the point $g$ it is
given by 
\be
R(t')=\frac{6+84kt'}{k(1+2kt')^2} \ . \label{Ricci}
\ee
The curvature thus increases at first, in agreement with the intuition that
the brane warps the space around it.

\noindent
The geometric picture breaks down as
soon as the curvature becomes too big. In fact, if one considers
$R(t')$ on all of $S^3$, it turns out that at $kt'=\frac{1}{2}$ the
curvature becomes singular at some points. The
geometric approximation thus becomes unreliable as soon as $kt'\sim
1$. In particular, one should not trust (\ref{Ricci}) for values
$kt'\sim\frac{5}{14}$, where $R(t')$ seemingly starts to decrease.


\section{Flat space}
\setcounter{equation}{0}
The last example we consider is the bosonic string in flat space
in the presence of a D$p$-brane. In this case, one
can consider the low-energy supergravity limit of the theory. The
D-brane is then given by a $p$-brane, a solution of the 
corresponding supergravity equation. \cite{DiVecchia:1997pr} performed
a boundary state calculation and found agreement with the supergravity
results. We will reproduce their results using the extended RG
equations.  

\subsection{The boundary state}
The conformal field theory is described by 26 free bosons with
ladder operators $a^\mu_n, \bar a^\nu_n$. 
A D$p$-brane located at $y$ is described by the boundary state
\cite{DiVecchia:1997pr} 
\be
||Dp;y\rangle\rangle = \frac{T_p}{2}\,\int
\frac{d^{d_\perp}k_\perp}{(2\pi)^{d_\perp}}  
e^{ik_\perp y}
\exp\left[-\sum_{n=1}^\infty a_{-n}^\mu\mathcal{S}_{\mu\nu}\bar
  a_{-n}^\nu\right] |0;k_\parallel=0,k_\perp\rangle\ \ .
\ee
The diagonal matrix $\mathcal{S}_{\mu\nu}$ is given by
\be
\mathcal{S}_{\mu\nu}=(\eta_{\alpha\beta},-\delta_{ij})\ ,
\ee
where $\alpha,\beta$ run over the $d_\parallel = p+1$ dimensions
parallel to the brane, and $i,j$ over the $d_\perp = 26-p-1$
transverse dimension.
Its tension is
\be
T_p=\frac{\sqrt{\pi}}{2^{(d-10)/4}}(4\pi^2\alpha')^{(d-2p-4)/4}\ .
\ee
Again, we will ignore the tachyon and concentrate
on marginal fields. The corresponding states are of the form
\be
a_{-1}^\mu \bar a_{-1}^\nu |0;k\rangle\  .\label{lvl1state}
\ee
Here $|0;k\rangle$ is the ground state of momentum $k$,
normalised as $\langle k|k'\rangle =
2\pi\delta(k-k')$, with 
$(2\pi)^d \delta^{(d)}(0)=V$.
The conformal weight of (\ref{lvl1state}) is
$(1+\alpha'k^2/4,1+\alpha'k^2/4)$, and it couples to
the D$p$-brane centred at $y=0$ as
\be
A^{\mu\nu}_k:= \langle 0;k|a_1^\mu \bar a_1^\nu||Dp;0\rangle\rangle =
-\frac{T_p}{2}\delta^{(d_{\parallel})}(k_\parallel)\mathcal{S}^{\mu\nu}
\ . \label{Amunu}
\ee
We see that only states with
$k_\parallel=0$ couple to the brane. It is thus necessary to consider
states with non-vanishing transverse momentum, which means
$k_\perp^2>0$, such that $k^2>0$. This poses a problem, as in string
theory vertex operators have to be marginal, so that $k^2=0$. 

\noindent This analysis indicates that we need to go off-shell
to find 
states that couple to the brane. From the CFT point of view such this
means that we need to consider states (\ref{lvl1state}) that are
almost marginal.

\subsection{Applying the RG equations}
We would like to apply (\ref{g1bulkflow}) and find the fixed
point to which the theory flows.
Although we derived (\ref{g1bulkflow}) only for
marginal fields, the argument also works for almost marginal fields with
$h=1+\delta h$. $\delta h$ then takes the role of $\epsilon$, and the
counterterm needed is $\sim 
\ell^{\delta h}(\delta h)^{-1}$. The contribution to
(\ref{g1bulkflow}) is again
$\frac{g_s}{\pi}\langle\phi_k||B\rangle\rangle$. 
It is clear however that several steps of the derivation depended on
taking $\epsilon 
\rightarrow 0$ in the end. We should therefore trust
(\ref{g1bulkflow}) only for almost marginal fields with $\delta h \ll 1$.

\noindent A fixed point of (\ref{g1bulkflow}) is given by
\be
0=\dot\lambda^{\mu\nu} = (2-h)\lambda^{\mu\nu} + \frac{g_s}{\pi}
A^{\mu\nu}_k  +
{\cal O}
(g_s\lambda,\lambda^3,g_s^2)\ ,
\ee
so that to lowest order 
\be
\lambda^{\mu\nu} = \frac{2g_s}{\pi\alpha'}
\frac{A^{\mu\nu}_k}{k_{\perp}^2}\ = \frac{g_s T_p
  (2\pi)^{p+1}V_{p+1}}{\pi\alpha'}
\frac{\mathcal{S}^{\mu\nu}}{k^2_\perp} \ . \label{lambda}
\ee
To compare to the metric in the supergravity solution, we calculate
the expectation value of the graviton, \ie its one-point function.
Assuming
that the fields $\phi_{\mu\nu}(k)$ were orthonormal in the original
theory, the perturbed one-point function of $a_{-1}^\mu
\bar a_{-1}^\nu |0;k\rangle$ is
\be
\langle \phi^{\mu\nu}(k)\rangle_\lambda=\lambda^{\sigma\rho}\langle
  \phi_{\sigma\rho}(k)\phi^{\mu\nu}(k) \rangle_0
  +O(\lambda^2)=\lambda^{\mu\nu} 
      + O(\lambda^2)\ . \label{phi1pt}
\ee
To obtain the expectation value of the graviton, we have to extract
the symmetric traceless part of (\ref{lambda}), as has been done in
\cite{DiVecchia:1997pr}. 
(\ref{lambda}) then agrees with their results,
up to a constant factor due to different normalisations.

\noindent Our analysis is only valid if $\alpha' k_\perp^2 \ll 
1$, since else $\delta h$ is too big. Moreover 
$\frac{g_sT_p}{\alpha ' k_\perp^2} \ll 1$ is needed, 
since otherwise higher order terms will become important.
Geometrically this means that we can only
consider weakly curved configurations, and only probe the long
distance limit. Our analysis is thus valid in the same range of
parameters as the supergravity calculation.


\section{Conclusions}
We have calculated the backreaction of a brane on the bulk
theory. The RG equations so obtained are a concrete realisation of the
Fischler-Susskind mechanism. For the free boson on a circle and for
the $SU(2)$ WZW-model, the 
resulting flows agree with geometric expectations. For flat space, we
are able to reproduce the long-distance behaviour of the supergravity
solution. 

\noindent An obvious extension of this work is to try to include higher
order terms in $g_s$. Technically, this is probably quite
challenging. There is however a more fundamental question: the
analysis of section~2 shows that annulus tadpoles can be compensated
by local counterterms, \ie that their effect can be expressed by
standard RG equations. It is not clear that this will also work for higher
order tadpoles, \eg for the disk with one thin handle that shrinks to
zero thickness.

\noindent The other natural extension is to generalise the RG
equations to theories with worldsheet supersymmetry.
This would allow to consider setups that are phenomenologically more
interesting. In particular, one could investigate
supersymmetric configurations similar to
\cite{Cvetic:2001nr}. In such a setup, shifting the closed string
moduli away from 
their original, supersymmetric values will 
induce a flow in the configuration of branes. These in turn will backreact
on the bulk. It would be interesting to find the end point of this
combined flow and  to check if the resulting theory is again
supersymmetric.

\vspace{1cm}

\centerline{\large \bf Acknowledgements}
\vskip .2cm
I want to thank my advisor Matthias Gaberdiel for suggesting
this topic to me, and for his constant support and advice
during the project itself.
I would also like to thank Costas Bachas, Ben Craps, Stefan Fredenhagen,
Ingo Kirsch, and Stefan Stieberger for helpful discussions.


\begin{thebibliography}{99}

\bibitem{Chaudhuri:1988qb}
S.~Chaudhuri and J.A.~Schwartz,
{\it A criterion for integrably marginal operators},
Phys.\ Lett.\ B {\bf 219} (1989) 291.

\bibitem{Recknagel:1998ih}
A.~Recknagel and V.~Schomerus,
{\it Boundary deformation theory and moduli spaces of D-branes},
Nucl.\ Phys.\ B {\bf 545} (1999) 233
{\tt [arXiv:hep-th/9811237]}.


\bibitem{Fredenhagen:2006dn}
  S.~Fredenhagen, M.R.~Gaberdiel and C.A.~Keller,
  {\it Bulk induced boundary perturbations},
  J.\ Phys.\ A {\bf 40} (2007) F17
  {\tt [arXiv:hep-th/0609034]}.



\bibitem{Fredenhagen:2007rx}
  S.~Fredenhagen, M.R.~Gaberdiel and C.A.~Keller,
  {\it Symmetries of perturbed conformal field theories},
  {\tt arXiv:0707.2511 [hep-th]}.


\bibitem{Cremmer:1973ig}
  E.~Cremmer and J.~Scherk,
  {\it Factorization of the pomeron sector and currents in the dual resonance
  model},
  Nucl.\ Phys.\  B {\bf 50} (1972) 222.

\bibitem{Clavelli:1973uk}
  L.~Clavelli and J.A.~Shapiro,
  {\it Pomeron factorization in general dual models},
  Nucl.\ Phys.\  B {\bf 57} (1973) 490.


\bibitem{Fischler:1986ci}
  W.~Fischler and L.~Susskind,
  {\it Dilaton tadpoles, string condensates and scale invariance},
  Phys.\ Lett.\ B {\bf 171} (1986) 383.

\bibitem{Fischler:1986tb}
  W.~Fischler and L.~Susskind,
  {\it Dilaton tadpoles, string condensates and scale invariance. 2},
  Phys.\ Lett.\ B {\bf 173} (1986) 262.

\bibitem{Callan:1986bc}
  C.G.~Callan, C.~Lovelace, C.R.~Nappi and S.A.~Yost,
  {\it String loop corrections to beta functions},
  Nucl.\ Phys.\ B {\bf 288} (1987) 525.


\bibitem{Cardy}
J.L.~Cardy, {\it Conformal invariance and statistical mechanics}, 
in: `Fields, strings and critical phenomena', proceedings of the Les
Houches summer school 1988, North Holland (1990).



\bibitem{Green:1987mn}
  M.B.~Green, J.H.~Schwarz and E.~Witten,
  {\it Superstring Theory. Vol. 2: Loop amplitudes, anomalies and
    phenomenology}, 
CUP. (1987).


\bibitem{Witten:1983ar}
  E.~Witten,
  {\it Nonabelian bosonization in two dimensions},
  Commun.\ Math.\ Phys.\  {\bf 92} (1984) 455.



\bibitem{Gaberdiel:2001xm}
M.R.~Gaberdiel, A.~Recknagel and G.M.T.~Watts,
{\it The conformal boundary states for SU(2) at level 1},
Nucl.\ Phys.\ B {\bf 626} (2002) 344
{\tt [arXiv:hep-th/0108102]}.

\bibitem{Callan:1994ub}
 C.G.~Callan, I.R.~Klebanov, A.W.W.~Ludwig and J.M.~Maldacena,
{\it Exact solution of a boundary conformal field theory},
Nucl.\ Phys.\ B {\bf 422} (1994) 417
{\tt [arXiv: hep-th/9402113]}.

\bibitem{Polchinski:1994my}
J.~Polchinski and L.~Thorlacius,
{\it Free fermion representation of a boundary conformal field
theory}, 
Phys.\ Rev.\ D {\bf 50} (1994) 622
{\tt [arXiv:hep-th/9404008]}.

\bibitem{FZ}
I.B.~Frenkel and Y.~Zhu, 
{\it Vertex operator algebras associated to representations of affine 
and Virasoro algebras}, 
Duke Math.\ J.\ {\bf 66} (1992) 123.

\bibitem{Gaberdiel:1998fs}
M.R.~Gaberdiel and P.~Goddard,
{\it Axiomatic conformal field theory},
Commun.\ Math.\ Phys.\  {\bf 209} (2000) 549
{\tt [arXiv:hep-th/9810019]}.



\bibitem{Hassan:1992gi}
  S.F.~Hassan and A.~Sen,
  {\it Marginal deformations of WZNW and coset models from O(D,D)
  transformation},
  Nucl.\ Phys.\ B {\bf 405} (1993) 143
  {\tt [arXiv:hep-th/9210121]}.

\bibitem{Alekseev:1998mc}
  A.Y.~Alekseev and V.~Schomerus,
  {\it D-branes in the WZW model},
  Phys.\ Rev.\  D {\bf 60} (1999) 061901
  {\tt [arXiv:hep-th/9812193]}.



\bibitem{Forste:2001gn}
  S.~F\"orste,
  {\it D-branes on a deformation of SU(2)},
  JHEP {\bf 0202} (2002) 022
  {\tt [arXiv: hep-th/0112193]}.

\bibitem{DiVecchia:1997pr}
  P.~Di Vecchia, M.~Frau, I.~Pesando, S.~Sciuto, A.~Lerda and R.~Russo,
  {\it Classical p-branes from boundary state},
  Nucl.\ Phys.\  B {\bf 507} (1997) 259
  {\tt [arXiv:hep-th/9707068]}.


\bibitem{Cvetic:2001nr}
  M.~Cvetic, G.~Shiu and A.M.~Uranga,
  {\it Chiral four-dimensional N = 1 supersymmetric type IIA orientifolds from
  intersecting D6-branes},
  Nucl.\ Phys.\  B {\bf 615} (2001) 3
  {\tt[arXiv:hep-th/0107166]}.




\end{thebibliography}
\end{document}